\documentclass[showpacs,groupedaddress,pra,letterpaper,twocolumn]{revtex4}
\usepackage{graphicx}

\input{tcilatex}

\begin{document}

\title{ Macroscopical Entangled Coherent State Generator in V configuration
atom system}
\author{Ling Zhou$^{1}$, Han Xiong$^{2}$}
\affiliation{$^{1}$School of Physics and Optoelectronic Technology, Dalian University of
Technology, Dalian 116024, P.R. China\\
$^{2}$Institute for Quantum Studies and Department of Physics, Texas A\&M
University, College Station, Texas 77843, USA\\
}

\begin{abstract}
In this paper, we propose a scheme to produce pure and macroscopical
entangled coherent state. When a three-level ''V'' configuration atom
interacts with a doubly reasonant cavity, under the strong classical driven
condition, entangled coherent state can be generated from vacuum fields. An
analytical solution for this system under the presence of cavity losses is
also given.
\end{abstract}

\maketitle

\section{ \ Introduction}

The generation of Schr\"{o}dinger cat states and entangled coherent states
serves the first step to use coherent states in quantum information
processes such as quantum sweeping and quantum teleportation\cite{an}.
Numerous schemes have been proposed to generate such an entangled coherent
state\cite{kim,zou,wang,brun,monroe,davidovich,solano,scully}. A scheme
based on a double electromagnetically induced transparency system has been
proposed in reference\cite{kim}. As for ion trap systems, the vibrational
motion of ions\cite{zou} and the entanglement swapping\cite{wang} seems
promising to generate an entangled coherent state.Using Kerr nonlinearity
and a 50/50 beam splitter,multidimensional entangled coherent states can be
generated on the condition that the initial state is a coherent state and
the interaction times are within certain range\cite{van}.

Cavity quantum electrodynamics (QED) has been shown to be a convenient
environment to generate both Schr\"{o}dinger cat states\cite{brun,monroe}
and entangled coherent states\cite{davidovich} in early days. Recently,
Solano \textit{et. al.} proposed a scheme to entangle two cavity modes
through the interaction of the cavity modes with a system of $N$ two-level
atoms inside the cavity\cite{solano}. In their scheme, the two cavity modes
interact with the same atomic transition and will put some restrictions on
these two cavity modes.

On the other hand, atomic coherence, which results from the coherent
superposition of different quantum states, can lead to many novel phenomena.
These include correlated spontaneous emission laser (CEL)\cite{scully},
lasing without inversion\cite{scully2} and electromagnetically induced
transparence\cite{HARRIS} \textit{etc.}. It has been known that two cavity
modes can be entangled when they interact with three-level atoms \cite%
{scully, agwarl} and atomic coherence plays an essential role in this
entanglement generation\cite{han, fuli}. Ref.\cite{han} showed that the
two-mode macroscopically entangled continuous-variable state could be
created in a CEL system where the atomic coherence was induced by a
classical driving field. Ref.\cite{fuli} studied the interaction of a ``V''\
type three-level atom and two thermal modes of a doubly resonant cavity and
showed that given a small amount of atomic coherence, entanglement could be
generated between these two thermal modes even when they initially had
arbitrarily high temperatures. However, the question whether an entangled
coherent state of two cavity modes can be generated through the interaction
with a single three-level atom has still not been answered there. Most
recently, our group have investigated that under large detuning condition, a
``$\lambda $'' three-level atom system interacting with two-mode field can
entangle the two-mode field\cite{mu}. The main drawback of the schemes \cite%
{van,mu} is that the initial coherent state are demand, and the interaction
time should be controled accurately otherwise we can not obtain entangled
coherent state.

In this paper, we propose a scheme to generate a macroscopic entangled
coherent state through the resonant interaction of two-mode field and a
three level ``$V$''\ (or ``$\lambda $'') configuration atom. Comparing this
scheme with Ref.\cite{van,mu}, we do not need prepare the initial coherent
state field state while the amplitude of entangled coherent state can be
amplified, which means the system can work as a entanglement generator.
Also, we do not need to control the interaction time accurately. It just
affect the amplitude of the entangled coherent state and have no effect on
the entanglement. Comparing our scheme with that of Refs\cite{solano}, the
two cavity modes in our scheme interact with different atomic transitions,
and thus can be easily manipulated.

\section{The theory and the scheme description\protect\bigskip}

\begin{figure}[h]
\includegraphics*[width=60mm,height=50mm]{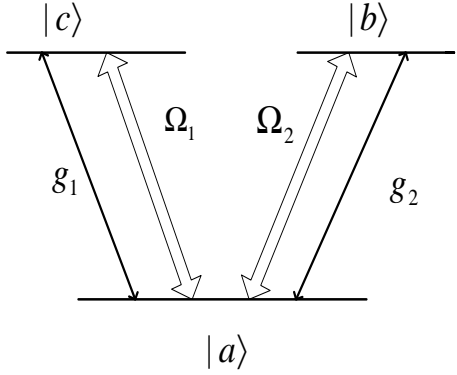}
\caption{The energy level configuration of the atom. Two cavity modes
resonantly interact with atomic transition $|c\rangle \leftrightarrow
|a\rangle $ and $|b\rangle \leftrightarrow |a\rangle $ while two classical
fields drive the two atomic transitions.}
\label{fig1}
\end{figure}

\bigskip We consider a three-level atom in the ``$V$ '' or ''$\lambda $''
configuration crossing a doubly resonant cavity. Here, we will take ''$V$''
configuration atom as example, and all of the calculation can be easy
generalized to ''$\lambda $'' configuration atom. The atomic level
configuration is depicted in Fig.\ref{fig1}. The atom resonantly interacts
with two cavity modes and two classical driving fields. $g_{1}$ and $g_{2}$
are two different atom-field coupling constants and $\Omega _{1}$, $\Omega
_{2}$ are the Rabi frequencies of the corresponding classical driving fields.

In the interaction picture, the interaction Hamiltonian has the following
form under the rotating-wave approximation

\begin{equation}
H=H_{0}+H_{1}  \label{eq1}
\end{equation}%
where 
\begin{eqnarray}
H_{0} &=&\Omega _{1}(|c\rangle \langle a|+|a\rangle \langle c|)  \nonumber
\label{eq2} \\
&&+\Omega _{2}(|b\rangle \langle a|+|a\rangle \langle b|),
\end{eqnarray}%
\begin{eqnarray}
H_{1} &=&g_{1}(a_{1}|c\rangle \langle a|+a_{1}^{+})|c\rangle \langle a|
\label{eq3} \\
&&+g_{2}(a_{2}|b\rangle \langle a|+a_{1}^{+})|a\rangle \langle b|.  \nonumber
\end{eqnarray}%
It is convenient to solve this system in a dressed state picture with
respect to the two strong classical driving fields. For this, we diagonalize 
$H_{0}$. The eigenstates and corresponding eigenvalues of $H_{0}$ are 
\begin{eqnarray}
|A\rangle &=&\frac{1}{u}(-\Omega _{1}|b\rangle +\Omega _{2}|c\rangle
),\lambda _{1}=0, \\
|B\rangle &=&\frac{1}{\sqrt{2}u}[u|a\rangle +\Omega _{2}|b\rangle +\Omega
_{1}|c\rangle ],\lambda _{2}=u,  \nonumber \\
|C\rangle &=&\frac{1}{\sqrt{2}u}[-u|a\rangle +\Omega _{2}|b\rangle +\Omega
_{1}|c\rangle ],\lambda _{3}=-u,  \nonumber
\end{eqnarray}%
where 
\[
u=\sqrt{\Omega _{1}^{2}+\Omega _{2}^{2}}. 
\]%
It is easy to prove that states $|A\rangle $, $|B\rangle $ and $|C\rangle $
compose a new orthogonal and complete basis of the three level system. Under
this basis, the atomic states can written as 
\begin{eqnarray}
|a\rangle &=&\frac{1}{\sqrt{2}}(|B\rangle -|C\rangle ),  \label{eq5} \\
|b\rangle &=&\frac{1}{\sqrt{2}u}[-\sqrt{2}\Omega _{1}|A\rangle +\Omega
_{2}(|B\rangle +|C\rangle )],  \nonumber \\
|c\rangle &=&\frac{1}{\sqrt{2}u}[\sqrt{2}\Omega _{2}|A\rangle +\Omega
_{1}(|B\rangle +|C\rangle )],  \nonumber
\end{eqnarray}%
We can then rewrite Hamiltonian Eq.(1) under this new basis set and perform
the following unitary transformation $H_{I}=e^{iH_{0}t}H_{1}e^{-iH_{0}t}$.
We have 
\begin{eqnarray}
H_{I} &=&\frac{1}{2u}\{g_{1}a_{1}[\Omega _{1}(|B\rangle \langle B|-|C\rangle
\langle C|  \label{eq6} \\
&&-|B\rangle \langle C|e^{-2uti}+|C\rangle \langle B|e^{2uti})  \nonumber \\
&&+\sqrt{2}\Omega _{2}(e^{uti}|A\rangle \langle B|-|A\rangle \langle
C|e^{-uti})]  \nonumber \\
&&+g_{2}a_{2}[\Omega _{2}|B\rangle \langle B|-|C\rangle \langle C|  \nonumber
\\
&&-|B\rangle \langle C|e^{-2uti}+|C\rangle \langle B|e^{2uti}  \nonumber \\
&&-\sqrt{2}\Omega _{1}(e^{uti}|A\rangle \langle B|-|A\rangle \langle
C|e^{-uti})]+h.c.\}  \nonumber
\end{eqnarray}%
In strong driving regime, that is $\Omega _{1}(\Omega _{2})\gg g_{1}(g_{2})$%
, we can realize a secular approximation (i.e. rotating-wave approximation)
and eliminate the high frequency terms in Eq.(6)\cite{solano}. The effective
Hamiltonian under this approximation is 
\begin{equation}
H_{eff}=\frac{1}{2u}[\Omega _{1}g_{1}(a_{1}+a_{1}^{+})+\Omega
_{2}g_{2}(a_{2}+g_{2}a_{2}^{+})](|B\rangle \langle B|-|C\rangle \langle C|).
\label{eq7}
\end{equation}%
If our initial state of the atom field combined system is $|\Psi (0)\rangle
=|a,0,0\rangle $. By using Hamiltonian Eq.(\ref{eq7}), we can have the state
of the system as 
\begin{equation}
|\Psi (t)\rangle =\frac{1}{\sqrt{2}}(|B,\alpha ,\beta \rangle -|C,-\alpha
,-\beta \rangle )  \label{eq8}
\end{equation}%
where $\alpha =\frac{-i\Omega _{1}g_{1}t}{2u}$ and $\beta =\frac{-i\Omega
_{2}g_{2}t}{2u}$. We now apply the inverse unitary transformation on state
Eq.(\ref{eq8})and change the basis set back to original atomic states, and
we have 
\begin{eqnarray}
|\Psi (t)\rangle &=&\frac{1}{\sqrt{2}}[\sqrt{2}|a\rangle (e^{-iut}|\alpha
,\beta \rangle +e^{iut}|-\alpha ,-\beta \rangle ) \\
&&+|b\rangle (e^{-iut}|\alpha ,\beta \rangle -e^{iut}|-\alpha ,-\beta
\rangle )  \nonumber \\
&&+|c\rangle (e^{-iut}|\alpha ,\beta \rangle -e^{iut}|-\alpha ,-\beta
\rangle )].  \nonumber
\end{eqnarray}%
When the atom comes out from the two mode cavity, we can use level-selective
ionizing counters to detect the atomic state. If the internal state of atom
is detected to be $|b\rangle ,|c\rangle $ or $|a\rangle $. The two-mode
field will be projected into 
\begin{equation}
|\Psi (t)\rangle _{f\pm }=\frac{1}{\sqrt{M_{\pm }}}[e^{-iut}|\alpha ,\beta
\rangle \pm e^{iut}|-\alpha ,-\beta \rangle ],  \label{eq10}
\end{equation}%
where%
\begin{equation}
M_{\pm }=2[1\pm \cos 2ut\exp (-2|\alpha |^{2}-2|\beta |^{2})].  \label{eq11}
\end{equation}%
The state Eq.(10) is a normalized one. The average photon number of the two
modes of the cavity can be easily obtained as 
\begin{eqnarray}
\langle N_{1}\rangle &=&\frac{2|\alpha |^{2}}{M}(1\mp e^{-2|\alpha
|^{2}-2|\beta |^{2}}\cos 2ut),  \label{eq11.1} \\
\langle N_{2}\rangle &=&\frac{2|\beta |^{2}}{M}(1\mp e^{-2|\alpha
|^{2}-2|\beta |^{2}}\cos 2ut).  \label{eq11.2}
\end{eqnarray}

We now try to estimate the entanglement of state Eq.(10). We notice that for
a general normalized and nonorthogonal entangled coherent state 
\begin{equation}
|\Psi \rangle =\mu |\bar{\alpha},\bar{\beta}\rangle +\nu |\bar{\gamma},\bar{%
\delta}\rangle ,  \label{eq12}
\end{equation}%
we can define $|0\rangle =|\bar{\alpha}\rangle $, $|1\rangle =(|\bar{\gamma}%
\rangle -p_{1}|\bar{\alpha}\rangle )/\sqrt{1-|p_{1}|^{2}}$ with $%
p_{1}=\langle \bar{\alpha}|\bar{\gamma}\rangle $ for the first subsystem and
define $|0\rangle =|\bar{\beta}\rangle $, $|1\rangle =(|\bar{\delta}\rangle
-p_{2}|\bar{\gamma}\rangle )/\sqrt{1-|p_{2}|^{2}}$ with $p_{2}=\langle \bar{%
\gamma}|\bar{\beta}\rangle $ for the second subsystem. The entanglement of
state Eq.(\ref{eq12}) can be measured on the orthogonal basis $|0,0\rangle $%
, $|0,1\rangle $, $|1,0\rangle $ and $|1,1\rangle $ \cite{wang,xgwang}. We
recall that the concurrence of state can be used to estimate the
entanglement for such a state. The concurrence\cite{wooters} of a state is
defined as $C=\max (0,2\max {\lambda _{i}}-\sum_{i=1}^{4}\lambda _{i})$
where $\lambda _{i}$ is the square roots of the eigenvalues of the matrix $%
R=\rho (\sigma _{1}^{y}\otimes \sigma _{2}^{y})\rho ^{\ast }(\sigma
_{1}^{y}\otimes \sigma _{2}^{y})$ and $\sigma _{i}^{y}$ is Pauli matrices.
The concurrence of state Eq.(\ref{eq12}) is\cite{wang,xgwang} 
\begin{equation}
C=2|\mu \nu |\sqrt{(1-|p_{1}|^{2})(1-p_{2}^{2})}.  \label{eq13}
\end{equation}%
Whenever $C>0$, state Eq.(\ref{eq12}) is an entangled state. The concurrence
of the state generated from our system (Eq.\ref{eq10}) is 
\begin{equation}
C=\frac{\sqrt{[(1-\exp (-4|\alpha |^{2})][(1-\exp (-4|\beta |^{2})]}}{1\pm
\cos 2ut\exp (-2|\alpha |^{2}-2|\beta |^{2})}.  \label{eq14}
\end{equation}

\begin{figure}[tph]
\includegraphics*[width=80mm,height=55mm]{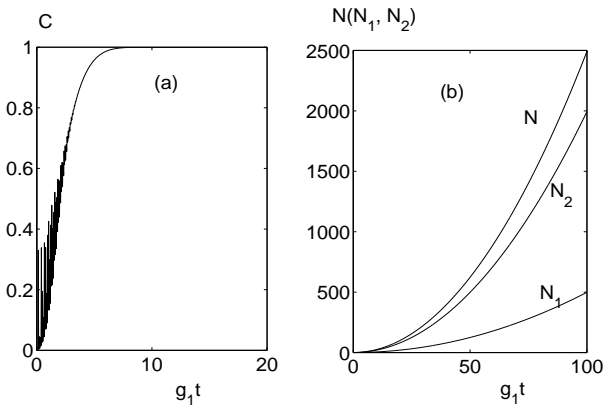}
\caption{The time evolution for entanglement and average photon number. In
Fig.2b, the lines labeled with $N$,$N_{1}$,$N_{2}$ correspond to total, mode
1, mode 2 average photon number, respectively. The parameters are $%
g_{1}=g_{2}=1$, $\Omega _{1}=100$, $\Omega _{2}=200.$}
\label{fig2}
\end{figure}

Fig.\ref{fig2} shows the time evolution of concurrence (Eq.(\ref{eq14})) and
the average total photon number (Eqs.(\ref{eq11.1}) and (\ref{eq11.2})) of
our state. Here, the positive sign has been chosen for Eq.(\ref{eq14}) and
negative sign has been chosen for Eqs.(\ref{eq11.1}) and (\ref{eq11.2}). As
we have analyzed in \cite{zhoul}, due to the phase $e^{\pm iut}$ , the state 
$|\Psi (t)\rangle _{f+}$ and $|\Psi (t)\rangle _{f-}$ almost have no
difference. One can see it from the expression of Eq.(15). The high
frequency term related with cos$2ut$ actually can be ignored. The property
is different from the two state $\frac{1}{\sqrt{N}}(|\alpha ,\alpha \rangle
\pm |-\alpha ,-\alpha \rangle $ \cite{WANG3}, where $\frac{1}{N_{-}}(|\alpha
,\alpha \rangle -|-\alpha ,-\alpha \rangle )$ is exact one ebit and its
entanglement is always 1 while $\frac{1}{N_{+}}(|\alpha ,\alpha \rangle
+|-\alpha ,-\alpha \rangle )$ is maximum state only when $\alpha \rightarrow
\infty $. From the Fig.2, we see that during short time evolution
concurrence curve show high frequency oscillation which come from classical
field. As time elapse, entanglement and average photon number increase. The
entanglement reaches its maximum value 1 after a specific time and in the
mean time we can obtain large number of photons in the cavity. This can be
understood clearly from Eq.(\ref{eq14}). If $\alpha $ and $\beta $ are large
enough, $\alpha $ and $-\alpha ,\beta $ and $-\beta $ will be orthogonal,
i.e., $\langle \alpha |-\alpha \rangle =\exp (-2|\alpha |^{2})=0$, $\langle
\beta |-\beta \rangle =\exp (-2|\beta |^{2})=0$ so that the state equal to $%
\frac{1}{\sqrt{2}}(|0,0\rangle +|1,1\rangle )$. Therefore concurrence $C=1$.
One also see that $\alpha $ and $\beta $ with increase with respect to time,
the average photon number is thus increased.

\section{The two-mode field in the leak cavities}

In order to obtain analytic solution of density matrix, here we do not
consider the atomic level decay. We now consider the effects of cavity
losses upon the entanglement generation of the two mode field. The master
equation is 
\begin{eqnarray}
\dot{\rho} &=&-i[H_{eff},\rho ]+\kappa (2a_{1}\rho
a_{1}^{+}-a_{1}^{+}a_{1}\rho -\rho a_{1}^{+}a_{1}  \nonumber  \label{eq2.1}
\\
&&+2a_{2}\rho a_{2}^{+}-a_{2}^{+}a_{2}\rho -\rho a_{2}^{+}a_{2})
\end{eqnarray}%
For simplicity, we assume the two modes have the same cavity decay rates $%
\kappa $. We still assume the atom is initially injected into state $%
|a\rangle $. Thus, we only need to work on the subspace $|B\rangle $ and $%
|C\rangle $. Using superoperator technique \cite{pei}and Hausdorff
similarity transformation \cite{wit}, we deduce that 
\begin{eqnarray}
\rho &=&|B,\alpha ^{\prime },\beta ^{\prime }\rangle \langle B,\alpha
^{\prime },\beta ^{\prime }|  \label{eq2.2} \\
&&+|C,-\alpha ^{\prime },-\beta ^{\prime }\rangle \langle C,-\alpha ^{\prime
},-\beta ^{\prime }|  \nonumber \\
&&-q(e^{-2uit}|B,\alpha ^{\prime },\beta ^{\prime }\rangle \langle C,-\alpha
^{\prime },-\beta ^{\prime }|  \nonumber \\
&&+e^{2uit}|C,-\alpha ^{\prime },-\beta ^{\prime }\rangle \langle B,\alpha
^{\prime },\beta ^{\prime }|)  \nonumber
\end{eqnarray}%
where 
\begin{eqnarray}
\alpha ^{\prime } &=&\frac{-i\Omega _{1}g_{1}}{2u\kappa }(1-e^{-\kappa
_{1}t}),\beta ^{\prime }=\frac{-i\Omega _{2}g_{2}}{2u\kappa }(1-e^{-\kappa
_{2}t}),  \nonumber \\
q &=&\exp [-\frac{\Omega _{1}^{2}g_{1}^{2}+\Omega _{2}^{2}g_{2}^{2}}{%
2u^{2}\kappa ^{2}}(2\kappa t+4e^{-\kappa t}-e^{-2\kappa t}-3)]  \label{eq2.3}
\end{eqnarray}%
with $P_{1}=\exp (-2|\alpha ^{\prime }|^{2})$, $P_{2}=\exp (-2|\beta
^{\prime }|^{2}$. In the process of calculation Eq.(19), we need
successively use the operator disentable equation $\exp (A+B)=\exp [\frac{%
A(1-e^{-\eta })}{\eta }]\exp (B)=\exp (B)\exp [\frac{A(e^{\eta }-1)}{\eta }]$
if $[A,B]=\eta A$. After the atom comes out from the cavity, we measure the
atomic state again. Suppose the internal state of the atom is detected to be
in $|a\rangle $, the two-mode field will be projected into 
\begin{eqnarray}
\rho &=&\frac{1}{S}[|\alpha ^{\prime },\beta ^{\prime }\rangle \langle
\alpha ^{\prime },\beta ^{\prime }|  \nonumber \\
&&+qe^{-2uit}|\alpha ^{\prime },\beta ^{\prime }\rangle \langle -\alpha
^{\prime },-\beta ^{\prime }|  \nonumber \\
&&+qe^{2uit}|-\alpha ^{\prime },-\beta ^{\prime }\rangle \langle \alpha
^{\prime },\beta ^{\prime }|  \nonumber \\
&&+|-\alpha ^{\prime },-\beta ^{\prime }\rangle \langle -\alpha ^{\prime
},-\beta ^{\prime }|]  \label{eq2.4}
\end{eqnarray}%
with 
\[
S=2+2qP_{1}P_{2}\cos 2ut. 
\]%
Now the field state is in a mixed entangled state. The difference between
the state Eq.(\ref{eq10}) and Eq.(\ref{eq2.4}) mainly lies in the factor of $%
q$ except for the change of coherent amplitude $\alpha ^{\prime }$. Here $q$
is the key factor to destroy the entanglement. With the time evolution, $q$
will achieve its asymptotic value zero so that the entanglement will be
destroyed completely. If there are no cavity losses, $q$ will be 1,
therefore the state Eq.(\ref{eq2.4}) will be exactly the same as Eq.(10).

To measure the entanglement, we will still use Concurrence. Let $|0\rangle
=|\alpha ^{\prime }\rangle $ , $|1\rangle =(|-\alpha ^{\prime }\rangle
-P_{1}|\alpha ^{\prime }\rangle )/M_{1}$ with $P_{1}=\langle \alpha ^{\prime
}|-\alpha ^{\prime }\rangle =\exp (-2|\alpha ^{\prime }|^{2})$, $M_{1}=\sqrt{%
1-|P_{1}|^{2}}$for field 1, and $|0\rangle =|\beta ^{\prime }\rangle $ , $%
|1\rangle =(|-\beta ^{\prime }\rangle -P_{2}|\beta ^{\prime }\rangle )/M_{2}$
with $P_{1}=\langle \beta ^{\prime }|-\beta ^{\prime }\rangle =\exp
(-2|\beta ^{\prime }|^{2})$, $M_{2}=$ $\sqrt{1-|P_{2}|^{2}}$for field 2. The
density matrix of the fields will be

\bigskip 
\begin{widetext}
\begin{equation}  \label{eq2.5}
\rho =\frac{1}{S}\left( 
\begin{array}{cccc}
1+P_{1}^{2}P_{2}^{2}+2qP_{1}P_{2}\cos 2ut & P_{1}M_{2}(P_{1}P_{2}+qe^{-2iut})
& P_{2}M_{1}(P_{1}P_{2}+qe^{-2iut}) & M_{1}M_{2}(P_{1}P_{2}+qe^{-2iut}) \\ 
P_{1}M_{2}(P_{1}P_{2}+qe^{2iut}) & P_{1}^{2}M_{2}^{2} & M_{1}M_{2}P_{1}P_{2}
& P_{1}M_{1}M_{2}^{2} \\ 
P_{2}M_{1}(P_{1}P_{2}+qe^{2iut}) & M_{1}M_{2}P_{1}P_{2} & P_{2}^{2}M_{1}^{2}
& P_{2}M_{1}^{2}M_{2} \\ 
M_{1}M_{2}(P_{1}P_{2}+qe^{2iut}) & P_{1}M_{1}M_{2}^{2} & M_{1}^{2}M_{2}P_{2}
& M_{1}^{2}M_{2}^{2}%
\end{array}%
\right)
\end{equation}%
\end{widetext}

Although the density matrix $\rho $ is very tedious, the eigenvalues of $R$
and concurrence still can be evaluated. The concurrence is 
\begin{equation}
C=\frac{2M_{1}M_{2}}{S}q.  \label{eq2.6}
\end{equation}%
One can easy check that without the loss of the cavity, the factor $q$ will
be 1 and the expression Eq.(22) will be the same as Eq.(16) for positive
sign. The average photon number for the two modes can be obtained as 
\begin{eqnarray}
N_{1} &=&\frac{2|\alpha ^{\prime }|^{2}}{S}(1-qP_{1}P_{2}\cos 2ut),
\label{eq2.7} \\
N_{2} &=&\frac{2|\beta ^{\prime }|^{2}}{S}(1-qP_{1}P_{2}\cos 2ut).  \nonumber
\end{eqnarray}%
\bigskip

\begin{figure}[tb]
\includegraphics*[width=80mm,height=55mm]{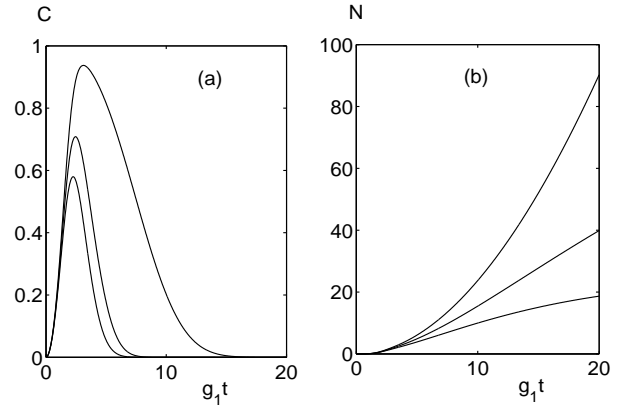}
\caption{The time evolution of entanglement and average photon number for
the loss of cavity $g_{1}=g_{2}=1$, $\Omega _{1}=200$, $\Omega _{2}=200.$%
From top to bottom for Fig.3a and 3b, $\protect\kappa =0.005$, 0.05, 0.1 ,
respectively.}
\label{fig3}
\end{figure}

Fig.\ref{fig3} shows the time evolution of the concurrence and the total
average photon number $N=N_{1}+N_{2}$ of the two-mode cavity under the
presence of cavity losses. Not surprisingly the average photon number of the
two-mode field will decrease with the increasing of $\kappa $. The
concurrence increases first and then drops down to zero. The reason of
concurrence dropping come from the factor $q$. With time going, $q$ will be
smaller and smaller so that entanglement is destroyed. Therefore, in our
scheme, high-Q doubly resonant cavity is preferred.

\section{Discussion and conclusion}

We now briefly address the experimental feasibility of the proposed scheme.
The required atomic level configuration can be achieved in Rydberg atoms
circular atomic level\cite{detection1}. The radiative lifetime is about $%
3\times 10^{-2}s$---are much longer than those for noncircular Rydberg states%
\cite{raimond}. Even in free space, the atoms would propagate a few meters
at thermal velocity before decaying. So, atomic radiative decay is thus
negligible along the 20-cm path inside the apparatus (In our model, we do
not consider the atomic spontaneous decay, therefore circular level Rydberg
atom is a good choice). For circular Rydberg atom, the coupling constant is $%
g=2\pi \times 24kHz$, if the interacting time $t\simeq 6.6\times 10^{-4}s$
(much smaller than the radiative lifetime $3\times 10^{-2}s$), $gt\sim 100$
correspond to the maximum value in Fig.2b, the coherent state $\alpha =\frac{%
-i\Omega _{1}g_{1}t}{2u}\sim -35i$ (for $\Omega _{1}=\Omega _{2})$.
Considering the loss of the cavity,$\alpha ^{\prime }=\frac{-i\Omega
_{1}g_{1}}{2u\kappa }(1-e^{-\kappa _{1}t})\sim -15.6i$ (for $\Omega
_{1}=\Omega _{2},\kappa =0.05)$. Therefore, based on a cavity QED techniques
presently, the proposed scheme might be realizable. As to the atomic
projecting detection, if the upper two atomic levels $|c\rangle $ and $%
|b\rangle $ are not degenerate, when the atom comes out from cavity, the
three field-ionization detector $D_{c},$ $D_{b}$ and $D_{a}$ can be used for
the three levels respectively. If the atomic lifetime is not long enough,
one must include the decay of the excited atomic level. The analytic
solution can not be obtained. One can numerically solve it. Definitely, the
decay of the atom will be bad for coherence of the system. In order to
obtain strong entanglement, high-Q doubly-resonant cavity and long lifetime
atom should be a first choice.

In conclusion, by employing a three-level ''V'' configuration atom
interacting with a two-mode cavity field, under the strong driven condition,
we can produce entangled coherent state from vacuum state, which means the
system can work as entanglement generator. The average photon number of the
two-mode cavity can be large. The scheme has its advantages. we do not need
to control the interacting time accurately. It just affect the amplitude of
the entanglement and have no effect on the entanglement. The two cavity
modes interact with different atomic transition so that it is easy to driven
two classical fields separately. The produced entangled coherent state can
be easily differentiated just by differentiable polarization.

Acknowledgments: Authors acknowledge the helpful discussion with Prof.
M.Suhail Zubairy and Dr. Yuri V. Rostovtsev. The project was supported by
NSFC under Grant No.10774020.


\begin{thebibliography}{99}
\bibitem{an} An N B 2004 Phys. Rev. A 69 022315; Jeong H, Kim M S, and Lee
Jinhyoung 2001 Phys. Rev. A 64 052308; van Enk S J and Hirota O 2001Phys.
Rev. A 64 022313

\bibitem{kim} Paternostro M, Kim M S, and Ham B S 2003 Phys. Rev. A 67
023811; Petrosyan D and Kurizki G 2002 Phys. Rev. A 65 33833

\bibitem{zou} Zou X B, Pahlke K, and Mathis W 2002 Phys. Rev. A 65 064303;
Gerry C C 1997 Phys. Rev. A 55 2478

\bibitem{wang} Wang X and Sanders B C 2001 Phys. Rev.A 65 012303

\bibitem{van} van Enk S J 2003 Phys. Rev. Lett. 91 017902

\bibitem{brun} Brune M, Hagley E, Dreyer J, Maitre X, Maali A, Wunderlich C,
Raimond J M, and Haroche S 1996 Phys. Rev. Lett. 77 4887

\bibitem{monroe} Monroe C, Meekhof D M, King B E, and Wineland D J 1996
Science 272 1131

\bibitem{davidovich} Davidovich L, Brune M, Raimond J M, and Haroche S 1996
Phys. Rev.A 53 1295

\bibitem{solano} Solano E, Agarwal G S and Walther H 2003 Phys.Rev.Lett. 90
027903

\bibitem{scully} Scully M O 1985 Phys. Rev. Lett. 55 2802; Scully M O,
Zubairy M S 1987 Phys. Rev.A 35 752

\bibitem{scully2} Scully M O, Zhu S Y, Gavrielides A 1989 Phys. Rev. Lett.
62 2813

\bibitem{HARRIS} Harris S E, Field J E, Imamoglu A 1990 Phys. Rev. Lett. 64
1107

\bibitem{agwarl} Agarwal G S 1993 Phys. Rev. Lett. 71 1351

\bibitem{han} Xiong H, Scully M O, and Zubairy M S 2005 Phys.Rev.Lett. 94
023601

\bibitem{fuli} Li F L, Xiong H, and Zubairy M S 2005 Phys. Rev. A 72
010303(R)

\bibitem{mu} Mu Q X, Ma Y H, Zhou L 2007 J. Phys. B : At. Mol. Opt. Phys. 40
3241

\bibitem{xgwang} Wang X 2002 J. Phys. A 35(1) 165

\bibitem{wooters} Hill S and Wootters W K 1997 Phys. Rev. Lett. 78 5022

\bibitem{zhoul} Zhou L, Yang G H 2006 J. Phys. B : At. Mol. Opt. Phys. 39
5143

\bibitem{WANG3} Wang X 2001Phys. Rev. A 64 022302

\bibitem{pei} Peixoto de Faria J G and Nemes M C 1999 Phys. Rev. A 59 3918

\bibitem{wit} Witschel W 1981 Int. J. Quantum Chem. 20 1233

\bibitem{detection1} Garcia-Maraver R, Corbalan R, Eckert K, Rebic S, Artoni
M, and Mompart J 2004 Phys. Rev. A 70 062324

\bibitem{raimond} Raimond J M, Brune M, and Haroche S 2001 Rev. Mod. Phys.
73 565
\end{thebibliography}
\end{document}